\documentclass[aps,twocolumn,superscriptaddress]{revtex4}
\usepackage{times}
\usepackage{color}
\usepackage{float}
\usepackage{amssymb}
\usepackage{epsfig}
\usepackage{mathrsfs}
\usepackage{amsmath}

\begin{document}

\title{Path integral approach to the calculation of the characteristic function of work}

\author{Tian Qiu}
\affiliation{Institute of Condensed Matter and Material Physics, School of Physics, Peking University, Beijing, 100871, China}

\author{Zhaoyu Fei}
\affiliation{Institute of Condensed Matter and Material Physics, School of Physics, Peking University, Beijing, 100871, China}

\author{Rui Pan}
\affiliation{Institute of Condensed Matter and Material Physics, School of Physics, Peking University, Beijing, 100871, China}

\author{H. T. Quan}\thanks{Corresponding author: htquan@pku.edu.cn}
\affiliation{Institute of Condensed Matter and Material Physics, School of Physics, Peking University, Beijing, 100871, China}
\affiliation{Collaborative Innovation Center of Quantum Matter, Beijing 100871, China}

\date{\today}

\begin{abstract}
Work statistics characterizes important features of a non-equilibrium thermodynamic process. But the calculation of the work statistics in an arbitrary non-equilibrium process is usually a cumbersome task. In this work, we study the work statistics in quantum systems by employing Feynman's path-integral approach. We derive the analytical work distributions of two prototype quantum systems. The results are proved to be equivalent to the results obtained based on Schr\"{o}dinger's formalism. We also calculate the work distributions in their classical counterparts by employing the path-integral approach. Our study demonstrates the effectiveness of the path-integral approach to the calculation of work statistics in both quantum and classical thermodynamics, and brings important insights to the understanding of the trajectory work in quantum systems.
\end{abstract}

\maketitle
\section{Introduction}
In the past few decades, there's been growing interests in microscopic systems down to the nanoscale where fluctuations dominate. As an extension of the traditional thermodynamics, a comprehensive framework known as stochastic thermodynamics was proposed \cite{Sekimoto2010, Seifert2012}. In this framework, work, heat and entropy production are defined as trajectory functionals \cite{Sekimoto2010, Seifert2012, Jarzynski1997a, Jarznyski2011, Klages2013}. As a consequence, the first law of thermodynamics is reformulated from ensemble level to individual trajectory level, and the second law is refined from inequalities to equalities \cite{Jarzynski1997a, Jarzynski1997b, Crooks1999, Crooks2000, Hummer2001}, which have been verified by enormous experiments \cite{Liphardt2002, Collin2005, Wang2002, Ciliberto2013, Pekola2019, Alemany2009, Blickle2006, An2015, Hoang2018}.

Although trajectory work is well-defined in classical stochastic thermodynamics, it is elusive in quantum thermodynamics, because work is not an observable (it characterizes a process rather than an instantaneous state of a system) \cite{Talkner2007}. There are numerous proposals for the definition of quantum work \cite{Talkner2007,Tasaki2000, Kurchan2001, Talkner2016, Yadalam2019, Sampaio2016, Brandner2016, M2017, KenFuno2018, Kwon2018, Liu2018, Suomela2015, Engel2007, Subasi2012, Hekking2013, Solinas2015, Baumer2018, Sampaio2018, Guarnieri2019, Strasberg2019, Venkatesh2015, Allahverdyan2014, Miller2017}, and the most widely accepted one is based on the so-called two-point measurement (TPM). According to this definition, two projective measurements over the instantaneous Hamiltonian are performed at the beginning and the end of the force protocol. The work in a single realization is determined by the difference of the two energy eigenvalues \cite{Tasaki2000, Kurchan2001, Talkner2007, Talkner2016}. It is straightforward to demonstrate that the TPM scheme leads to a quantum version of the Jarzynski equality \cite{Jarzynski1997a} and the Crooks fluctuation theorem \cite{Crooks1999}.

By analogy with the partition function, which characterizes completely the properties of a thermal equilibrium state, the work statistics is an essential function which encodes important information about the non-equilibrium thermodynamic process \cite{Dora2012, Goold2018}. Nevertheless, the calculation of the work statistics in an arbitrary non-equilibrium process is a cumbersome task, and there is no universal method to do the calculation. In literature, the work statistics are usually calculated case by case, e.g., by solving the time-dependent Schr\"{o}dinger equation \cite{Deffner2008, Deffner2010, Quan2011, Gong2014} or Heisenberg equation \cite{Talkner2008, Fei2019}. In Refs. \cite{Liu2012, Fei2018}, a quantum Feynman-Kac equation was introduced, and the work statistics can be obtained by solving this equation. In this article, we propose a universal method, i.e., the path-integral approach to study the calculation of the work statistics in non-equilibrium processes. As examples, we calculate the work statistics in two prototype models, i.e., the harmonic oscillator with a time-dependent angular frequency and a free particle inside an expanding piston. We derive the analytical work distributions of non-equilibrium processes for these two systems. Our work provides good examples to show the effectiveness of the path-integral approach, and will shed new light on the understanding of quantum work.

This paper is organized as follows. In Sec. II, we briefly recall the path integral expression of the characteristic function of work and introduce two prototype models. In Sec. III, we derive the analytical work distributions of these two models by the path-integral approach and show the consistency with previous results. In Sec. IV, we do the calculation by employing the path-integral approach in their classical counterparts. We conclude our paper in Sec. V.

\section{Models and path integral expression of the characteristic function of work}
We consider an isolated system, the Hamiltonian is given by
\begin{equation}\label{eq:Hamiltonian}
\hat{H}(\lambda_t)=\frac{\hat{p}^2}{2m}+\hat{V}(\lambda_t,\hat{x}),
\end{equation}
where ${m}$ is the mass and ${\hat{V}(\lambda_t,\hat{x})}$ is an arbitrary potential with its time dependence specified by the protocal ${\lambda_t}$. Work is done on the system when the work parameter ${\lambda_t}$ is controlled by an external agent. We measure the energy of the system at ${t=0}$ and ${t=\tau}$ respectively, and get instantaneous eigenenergies ${E^0_n}$ (at ${t=0}$) and ${E^{\tau}_l}$ (at ${t=\tau}$). Then the quantum fluctuating work is defined as the difference between the two eigenenergies:
\begin{equation}\label{eq:work}
W_{l,n}=E^{\tau}_l-E^0_n.
\end{equation}
We use ${| E^t_n \rangle}$ to represent the ${n}$-th instantaneous energy eigenstate of the system at time ${t}$, and
\begin{equation}\label{eq:PN}
p_n=\left\langle E^0_n \left| \hat{\rho}(0) \right| E^0_n \right\rangle
\end{equation}
is the probability of finding the system in the ${n}$-th eigenstate ${| E^0_n \rangle}$ in the first projective measurement, where
\begin{equation}\label{eq:DM1}
\hat{\rho}(0)=e^{-\beta \hat{H}(\lambda_0)} / Z_0
\end{equation}
is the initial canonical density matrix of the system at the inverse temperature ${\beta}$, and ${Z_0=Tr[e^{-\beta \hat{H}(\lambda_0)}]}$ is the initial partition function. Then the joint probability of observing the fluctuating work ${W_{l,n}}$ is given by
\begin{equation}\label{eq:JP}
p(n,l)=p_n{\big| \langle E^\tau_l | \hat{U} | E^0_n \rangle \big| }^2,
\end{equation}
where
\begin{equation}\label{eq:US}
\hat{U}=\hat{T} e^{-\frac{i}{\hbar} \int_{0}^{\tau}dt \hat{H}(\lambda_t)}
\end{equation}
is the unitary operator of the evolution. The work probability distribution is given by
\begin{equation}\label{eq:WPD}
P(W)=\sum_{l,n} \delta(W-W_{l,n})\ p(l,n).
\end{equation}
Taking the Fourier transformation of the work probability distribution, we obtain the characteristic function of work \cite{Talkner2007}
\begin{equation}\label{eq:characteristic function of work}
\chi_W(\nu)=\int dW P(W) e^{i\nu W}.
\end{equation}
This can be written as
\begin{equation}\label{eq:CFW1}
\chi_W(\nu)=Tr\left[ \hat{U} e^{-i \nu \hat{H}(\lambda_0)} \hat{\rho}(0) \hat{U}^\dagger e^{i\nu \hat{H}(\lambda_\tau)} \right].
\end{equation}

Considering the definition of a propagator, we have the following relations: ${\langle x_f | \hat{U} e^{-i \nu \hat{H}(\lambda_0)} | x_i \rangle=\int Dx\ e^{\frac{i}{\hbar}S_1[x]}}$ and ${\langle y_i | \hat{U}^\dagger e^{i\nu \hat{H}(\lambda_\tau)} | y_f \rangle=\int Dy\ e^{-\frac{i}{\hbar}S_2[y]}}$, where ${Dx}$ and ${Dy}$ denote the integral over path ${x}$ and ${y}$, ${S_1[x]}$ and ${S_2[y]}$ are actions of the forward and the backward propagations which are defined as
\begin{subequations}\label{eq:S1S2}
\begin{equation}\label{eq:S1S2a}
S_1\!\left[x\right]\!=\!\int^{\hbar\nu}_{0}\!dt\ \mathscr{L}\!\left[  \lambda_0,\!x(t) \right]\!+\!
\int^{\tau\!+\!\hbar\nu}_{\hbar\nu}\!dt\ \mathscr{L}\!\left[  \lambda_{t\!-\!\hbar\nu},\!x(t) \right],
\end{equation}
\begin{equation}\label{eq:S1S2b}
S_2\!\left[y\right]\!=\!\int^{\tau}_{0}\!ds\ \mathscr{L}\!\left[  \lambda_s,\!y(s) \right]\!+\!
\int^{\tau\!+\!\hbar\nu}_{\tau}\!ds\ \mathscr{L}\!\left[  \lambda_{\tau},\!y(s) \right].\ \ \ \ \
\end{equation}
\end{subequations}
Here ${\mathscr{L}\!\left[ \lambda_t,x(t) \right]\!=\!\frac{m}{2}\dot{x}^2(t)\!-\!V\left(\lambda_t,x(t)\right)}$ is the Lagrangian. As a result, we can rewrite Eq. (\ref{eq:CFW1}) in the path integral formalism as \cite{Ken2018}
\begin{equation}\label{eq:CFW2}
\chi_W(\nu)=\int e^{\frac{i}{\hbar}\left(S_1[x]-S_2[y]\right)} \rho(x_i,y_i) \delta(x_f-y_f),
\end{equation}
where
\begin{equation}\label{eq:DM2}
\rho(x_i,y_i)=\left\langle x_i \left| \rho(0) \right| y_i \right\rangle
\end{equation}
is the elements of the density matrix in the coordinate representation, and the integration in Eq. (\ref{eq:CFW2}) is ${\int dx_i dx_f dy_i dy_f Dx Dy}$.

One can see that there are two different propagations in Eq. (\ref{eq:CFW2}), i.e., the forward and the backward propagations. For the forward propagation denoted by ${S_1[x]}$, the work parameter is held fixed at ${\lambda_0}$ during the period from ${t=0}$ to ${t=\hbar\nu}$, then the work parameter is changed from an initial value ${\lambda_0}$ at ${t=\hbar\nu}$ to a final value ${\lambda_\tau}$ at ${t=\tau+\hbar\nu}$. While for the backward propagation denoted by ${S_2[y]}$, the work parameter is first changed from the initial value ${\lambda_0}$ at ${s=0}$ to a final value ${\lambda_\tau}$ at ${s=\tau}$, then the work parameter is held fixed at its final value ${\lambda_\tau}$ during the period from ${s=\tau}$ to ${s=\tau+\hbar\nu}$.

As is known, the Feynman path integrals are usually difficult to evaluate. However, in the semiclassical approximation, only the classical paths are included in the path integral and the evaluation of path integral becomes much simpler. In some special cases, the semiclassical propagators are exact, for example, the two prototype models considered in the current paper. As the first example, we consider the quantum harmonic oscillator with a time-dependent angular frequency \cite{Husimi1953, Deffner2008, Deffner2010}, and the other one is a free particle inside a rigid box with one wall moving uniformly in time \cite{Luz1992}. We hope to use these two examples to illustrate the effectiveness of the path-integral approach to the calculation of the work statistics in non-equilibrium processes and also gain insights about the meaning of quantum trajectory work.

\section{Path-integral approach to the calculation of work statistics in two prototype quantum systems}
In the following, we will evaluate the characteristic function ${\chi_W(\nu)}$ (\ref{eq:CFW2}) of two prototype models. One can see that the forward (\ref{eq:S1S2a}) (also the backward (\ref{eq:S1S2b})) propagation can be divided into two independent parts by inserting an intermediate position ${x_b}$. For the forward propagation:
\begin{equation}\label{eq:Forward}
\int^{x_f}_{x_i}\!Dx\ e^{\frac{i}{\hbar}\!S_1\![x]}\!=\!
\int\!dx_b\int^{x_b}_{x_i}\!Dx\
e^{\frac{i}{\hbar}\!I_1\![x]}\!\int^{x_f}_{x_b}\!Dx\
e^{\frac{i}{\hbar}\!I_2\![x]},
\end{equation}
where
%\begin{eqnarray}\label{eq:I1I2}
%I_1[x]&=&\int^{\hbar\nu}_{0}dt\ \mathscr{L}\left[  \lambda_0,x(t) %\right], \nonumber \\
%I_2[x]&=&\int^{\tau}_{0}dt\ \mathscr{L}\left[  \lambda_t,x(t) \right].
%\end{eqnarray}
\begin{equation}\label{eq:I1I2}
I_1[x]\!=\!\int^{\hbar\nu}_{0}\!dt\ \mathscr{L}\!\left[  \lambda_0,x(t) \right],
I_2[x]\!=\!\int^{\tau}_{0}\!dt\ \mathscr{L}\!\left[  \lambda_t,x(t) \right].
\end{equation}
For the backward propagation:
\begin{equation}\label{eq:Backward}
\int^{y_f}_{y_i}\!Dy\ e^{\!-\!\frac{i}{\hbar}\!S_2\![y]}=
\int\!dy_b\int^{y_b}_{y_i}\!Dy\
e^{\!-\!\frac{i}{\hbar}\!I_3\![y]}
\int^{y_f}_{y_b}\!Dy\
e^{\!-\!\frac{i}{\hbar}\!I_4\![y]},
\end{equation}
where
\begin{equation}\label{eq:I3I4}
I_3[y]\!=\!\int^{\tau}_{0}\!ds\ \mathscr{L}\!\left[  \lambda_s,y(s) \right],
I_4[y]\!=\!\int^{\hbar\nu}_{0}\!ds\ \mathscr{L}\!\left[  \lambda_\tau,y(s) \right].
\end{equation}
We first calculate each part independently and then integrate over the intermediate position ${x_b}$ or ${y_b}$ to get the propagator of the forward and the backward propagation.

\subsection{Quantum harmonic oscillator with a time-dependent angular frequency}
We consider a quantum harmonic oscillator with a time-dependent angular frequency ${\omega(t)}$. Please note that ${\omega(t)}$ plays the role of ${\lambda_t}$ in this model and can be an arbitrary function of ${t}$ (not necessarily a linear function of ${t}$). The forward propagation consists of two parts. For the first part, the angular frequency of the quantum harmonic oscillator is fixed at ${\omega_0}$ during the period from ${t=0}$ to ${t=\hbar \nu}$. For the second part, the frequency is changed from the initial value ${\omega_0}$ at ${t=\hbar \nu}$ to ${\omega_1}$ at ${t=\hbar \nu+\tau}$ according to a given protocal ${\omega(t)}$. The backward propagation also consists of two parts. For the first part, the frequency is changed from ${\omega_0}$ at ${t=0}$ to ${\omega_1}$ at ${t=\tau}$ according to the same work protocal ${\omega(t)}$ as in the forward propagation. For the second part, the frequency of the system is fixed at ${\omega_1}$ during the period from ${t=\tau}$ to ${t=\tau+\hbar \nu}$.

In the following, we will derive the characteristic function of work based on the path integral approach. First, let us derive the semiclassical propagator of the system. The Lagrangian of the system can be written as

\begin{equation}\label{eq:Lagrangian2}
\mathscr{L}\left[ \lambda_t,x(t) \right]:=\frac{m}{2}\dot{x}^2(t)-\frac{m}{2}\omega^2(t)x^2(t),
\end{equation}
and the classical path ${x(t)}$ satisfies
\begin{equation}\label{eq:CP}
\frac{d}{dt}\frac{\partial\mathscr{L}}{\partial\dot{x}}=\frac{\partial\mathscr{L}}{\partial x},
\end{equation}
or alternatively,
\begin{equation}\label{eq:CP2}
\ddot{x}(t)=-\omega^2(t)x(t)
\end{equation}
with the boundary conditions ${x(t_i)=x_i}$ and ${\ x(t_f)=x_f}$. Please note that the quantum harmonic oscillator has only one classical path. After some calculations, we get  the action of the system as \cite{Husimi1953}
\begin{equation}\label{eq:S-HO}
S\left[x_f\right]
%\int^{t}_{t_0} dt\ \left[\frac{m}{2}\dot{x}^2-\frac{m}{2}\omega^2 x^2\right]
=\frac{m}{2X(t_f)}\left(\dot{X}(t_f) x_f^2-2 x_f x_i+Y(t_f) x^2_i\right).
\end{equation}
Here $X(t)$ and ${Y(t)}$ are two linearly independent solutions to the second-order ordinary differential equation (\ref{eq:CP2}) with the initial conditions ${X(t_i)=0}$, ${\dot{X}(t_i)=1}$ and ${Y(t_i)=1}$, ${\dot{Y}(t_i)=0}$. They satisfy the relation ${\dot{X}Y-\dot{Y}X=1}$ for any time ${t}$. Then we get the semiclassical propagator
\begin{widetext}
\begin{equation}\label{eq:U-HO}
\int^{x_f}_{x_i}\!Dx\ e^{\frac{i}{\hbar}\mathcal{S}[x]}=\sqrt{\frac{m}{2\pi i \hbar X(t_f)}}\exp{\left[\frac{im}{2\hbar X(t_f)}
\left(\dot{X}(t_f) x^2_f-2 x_f x_i+Y(t_f) x^2_i\right)\right]}.
\end{equation}
\end{widetext}
%which is exactly the same with the propagator resulting from the self-consistent wave function approach \cite{Husimi1953}.
%Just as the moving piston model, we can also prove that
This semiclassical propagator can be proved to be exact \cite{Husimi1953}.

From Eq. (\ref{eq:CFW2}), one can see that the expression of the characteristic function of work consists of three parts. The first part is the propagator for the forward propagation, the second part is the propagator for the backward propagation, and the third part is the elements of the initial density matrix of the system in the coordinate representation. In the following, we will calculate these three parts separately, and then integrate over the initial and the final positions ${(x_i, y_i, x_f, y_f)}$.

{\it Propagator for the forward propagation.}---For the first part of the forward propagation, the angular frequency is fixed at the initial value ${\omega_0}$, and the oscillator moves from ${x_i}$ at ${t_i=0}$ to ${x_b}$ at ${t_f=\hbar\nu}$. The analytical expression of the propagator for the first part of the forward propagation can be expressed as:
\begin{widetext}
\begin{equation}\label{eq:I1d-HO}
\int^{x_b}_{x_i} Dx\ e^{\frac{i}{\hbar}I_1[x]}=
\sqrt{\frac{m}{2\pi i \hbar X_1(\hbar\nu)}}\exp{\left[\frac{im}{2\hbar X_1(\hbar\nu)}
\left(\dot{X_1}(\hbar\nu) x^2_b-2 x_b x_i+Y_1(\hbar\nu) x^2_i\right)\right]}.
\end{equation}
\end{widetext}
Here ${X_1(t)}$ and ${Y_1(t)}$ satisfy ${\ddot{X}_1=-\omega^2_0 X_1}$, ${\ddot{Y}_1=-\omega^2_0 Y_1}$, with the initial conditions ${X_1(0)=0}$, ${\dot{X}_1(0)=1}$, ${Y_1(0)=1}$, ${\dot{Y}_1(0)=0}$. The exact solution of ${X_1(\hbar\nu)}$ and ${Y_1(\hbar\nu)}$ can be obtained as follows,
\begin{eqnarray}\label{eq:XY-HO2}
X_1(\hbar\nu)=\sin{\nu \hbar\omega_0}/\omega_0,\ \ \dot{X}_1(\hbar\nu)=\cos{\nu \hbar\omega_0},\nonumber \\
Y_1(\hbar\nu)=\cos{\nu \hbar\omega_0},\ \ \dot{Y}_1(\hbar\nu)=-\omega_0 \sin{\nu \hbar\omega_0}.
\end{eqnarray}

For the second part of the forward propagation, the frequency is changed from the initial value ${\omega_0}$ to the final value ${\omega_1}$, and the oscillator moves from ${x_b}$ at ${t_i=0}$ to ${x_f}$ at ${t_f=\tau}$. The analytical expression of the propagator for the second part of the forward propagation can be expressed as
\begin{widetext}
\begin{equation}\label{eq:I2d-HO}
\int^{x_f}_{x_b} Dx\ e^{\frac{i}{\hbar}I_2[x]}=
\sqrt{\frac{m}{2\pi i \hbar X(\tau)}}\exp{\left[\frac{im}{2\hbar X(\tau)}
\left(\dot{X}(\tau) x^2_f-2 x_f x_b+Y(\tau) x^2_b\right)\right]}.
\end{equation}
\end{widetext}
Substituting Eqs. (\ref{eq:I1d-HO}, \ref{eq:I2d-HO}) into Eq. (\ref{eq:Forward}) and integrating over the intermediate position ${x_b}$, one can obtain the semiclassical propagator for the forward propagation
\begin{widetext}
\begin{eqnarray}\label{eq:S1a-HO}
\int^{x_f}_{x_i} Dx\ e^{\frac{i}{\hbar}S_1[x]}
&=&\sqrt{\frac{\pi}{2}}(1+i)\frac{m}{2\pi i\hbar}
\frac{1}{\sqrt{X_1(\hbar\nu)X(\tau)}}
\frac{1}{\sqrt{\frac{m}{2\hbar}\left[\frac{\dot{X}_1(\hbar\nu)}{X_1(\hbar\nu)}+\frac{Y(\tau)}{X(\tau)}\right]}}\nonumber\\
&\times&\exp{\left[-\frac{im}{2\hbar}\frac{\left(\frac{x_i}{X_1(\hbar\nu)}+\frac{x_f}{X(\tau)}\right)^2}
{\left(\frac{\dot{X}_1(\hbar\nu)}{X_1(\hbar\nu)}+\frac{Y(\tau)}{X(\tau)}\right)}+
\frac{im}{2\hbar}\left(\frac{Y_1(\hbar\nu)}{X_1(\hbar\nu)}x^2_i+\frac{\dot{X}(\tau)}{X(\tau)}x^2_f\right)\right]}.
\end{eqnarray}
\end{widetext}
Here we have used the results of Fresnel integration \cite{Gradshteyn2007}
\begin{equation}\label{eq:FI}
\int^{\infty}_{-\infty} dx_b\ e^{iC_1 x^2_b-iC_2 x_b}=
\sqrt{\frac{\pi}{2}}(1+i)\frac{e^{-\frac{iC^2_2}{4C_1}}}{\sqrt{C_1}},
\end{equation}
where ${C_1}$ and ${C_2}$ are two arbitrary constants.
%\begin{equation}\label{eq:model}
%\int^{x_f}_{x_i} Dx\ e^{\frac{i}{\hbar}S_1[x]}
%=\frac{4}{l_0\sqrt{l_0 l_f}}\sum^{\infty}_{n_1,n_2=1}
%e^{-\frac{i n^2_1 \pi^2 \hbar^2\nu}{2m l^2_0}+
%\frac{imu x^2_f}{2\hbar l_f}
%+\frac{i n^2_2 \pi^2 \hbar}{2mu}\left(\frac{1}{l_f}-\frac{1}{l_0}\right)}
%sin\left(\frac{n_1\pi x_i}{l_0}\right)sin\left(\frac{n_2\pi x_f}{l_f}\right)
%\end{equation}
%\begin{equation}\label{eq:model}
%\times\int^{l_0}_{0} dx_b\ e^{-\frac{imu x^2_b}{2\hbar l_0}}
%sin\left(\frac{n_1\pi x_b}{l_0}\right)sin\left(\frac{n_2\pi x_b}{l_0}\right),
%\end{equation}
%\begin{equation}\label{eq:model}
%\lim_{\nu \to 0,\tau \to 0}\int^{l_0}_{0} dx_i \int^{x_f}_{x_i} Dx\ e^{\frac{i}{\hbar}S_1[x]}=1
%\end{equation}

{\it Propagator for the backward propagation.}---Similarly, one can obtain the expression of the propagator for the backward propagation,
\begin{widetext}
\begin{eqnarray}\label{eq:S2a-HO}
\int^{y_f}_{y_i} Dx\ e^{-\frac{i}{\hbar}S_2[y]}
&=&\sqrt{\frac{\pi}{2}}(1-i)\frac{m}{-2\pi i\hbar}
\frac{1}{\sqrt{X(\tau)X_2(\hbar\nu)}}
\frac{1}
{\sqrt{\frac{m}{2\hbar}\left[\frac{\dot{X}(\tau)}{X(\tau)}+\frac{Y_2(\hbar\nu)}{X_2(\hbar\nu)}\right]}}\nonumber \\
&\times&\exp{\left[\frac{im}{2\hbar}\frac{\left(\frac{y_i}{X(\tau)}+\frac{y_f}{X_2(\hbar\nu)}\right)^2}
{\left(\frac{\dot{X}(\tau)}{X(\tau)}+\frac{Y_2(\hbar\nu)}{X_2(\hbar\nu)}\right)}
-\frac{im}{2\hbar}\left(\frac{Y(\tau)}{X(\tau)}y^2_i+\frac{\dot{X}_2(\hbar\nu)}{X_2(\hbar\nu)}y^2_f\right)\right]}.
\end{eqnarray}
\end{widetext}
Here ${X_2(t)}$ and ${Y_2(t)}$ satisfy ${\ddot{X}_2=-\omega^2_1 X_2,\ \ \ddot{Y}_2=-\omega^2_1 Y_2}$, with the initial conditions ${X_2(0)=0}$, ${\dot{X}_2(0)=1}$, ${Y_2(0)=1}$, ${\dot{Y}_2(0)=0}$. Similarly, the exact solutions of ${X_2(\hbar\nu)}$ and ${Y_2(\hbar\nu)}$ can be obtained as follows,
\begin{eqnarray}\label{eq:XY2-HO2}
X_2(\hbar\nu)=\sin{\nu \hbar\omega_1}/\omega_1,\ \ \dot{X}_2(\hbar\nu)=\cos{\nu \hbar\omega_1},\nonumber \\
Y_2(\hbar\nu)=\cos{\nu \hbar\omega_1},\ \ \dot{Y}_2(\hbar\nu)=-\omega_1 \sin{\nu \hbar\omega_1}.
\end{eqnarray}
%in which ${\varepsilon_1=\hbar\omega_1}$.
%\begin{equation}\label{eq:model}
%\int^{y_f}_{y_i} Dy\ e^{-\frac{i}{\hbar}S_2[y]}
%=\frac{4}{l_f\sqrt{l_0 l_f}}\sum^{\infty}_{n_3,n_4=1}
%e^{\frac{i n^2_3 \pi^2 \hbar^2\nu}{2m l^2_f}+
%\frac{imu y^2_i}{2\hbar l_0}
%+\frac{i n^2_4 \pi^2 \hbar}{2mu}\left(\frac{1}{l_0}-\frac{1}{l_f}\right)}
%sin\left(\frac{n_4\pi y_i}{l_0}\right)sin\left(\frac{n_3\pi x_f}{l_f}\right)
%\end{equation}
%\begin{equation}\label{eq:model}
%\times\int^{l_f}_{0} dy_b\ e^{-\frac{imu y^2_b}{2\hbar l_f}}
%sin\left(\frac{n_3\pi y_b}{l_f}\right)sin\left(\frac{n_4\pi y_b}{l_f}\right)
%\end{equation}

{\it Initial density matrix in the coordinate representation.}---We assume that the system is initially prepared in a thermal equilibrium state,
%\begin{equation}\label{eq:DMa-HO}
%\rho_S(0)=\sum^{\infty}_{n=1} \frac{e^{-\beta E^0_{n}}}{Z_0}\left| n \right\rangle \left\langle n \right|,
%\end{equation}
%then
\begin{equation}\label{eq:DMb-HO}
\rho(x_i,y_i)
%=\sum^{\infty}_{n=1}\frac{e^{-\beta E^0_{n}}}{Z_0}
%\left\langle x_i\left| n \right\rangle \left\langle n \right|y_i\right\rangle
=\sum^{\infty}_{n=1}\frac{e^{-\beta E^0_{n}}}{Z_0} \Psi^0_{n}(x_i){\Psi^0_{n}}^*(y_i),
\end{equation}
where
${E^0_n=\hbar\omega_0(n+1/2)}$, ${Z_0={e^\frac{-\beta\hbar\omega_0}{2}}/(1-e^{\beta\hbar\omega_0})}$
are the eigenenergies and the partition function of the quantum harmonic oscillator at time ${t=0}$, and ${\Psi^0_{n}(x)}$ are the corresponding eigenfunctions. The density matrix in the coordinate representation can be written in the following compact form \cite{Deffner2008, Deffner2010},
\begin{widetext}
\begin{eqnarray}\label{eq:DMd-HO}
\rho(x_i,y_i)&=&\left(1-e^{-\beta\hbar\omega_0}\right)
\sum^{\infty}_{n=1}\left(e^{-\beta\hbar\omega_0}\right)^n \Psi^0_{n}(x_i){\Psi^0_{n}}^*(y_i) \nonumber \\
&=&\left(1-e^{-\beta\hbar\omega_0}\right)\sqrt{\frac{m\omega_0}{\hbar\pi\left(1-e^{-2\beta\hbar\omega_0}\right)}}
\exp{\left[-\frac{m\omega_0}{\hbar}\frac{\left(1+e^{-2\beta\hbar\omega_0}\right)
\left(x^2_i+y^2_i\right)-4e^{-\beta\hbar\omega_0}x_i y_i}{2\left(1-e^{-2\beta\hbar\omega_0}\right)}\right]}.
\end{eqnarray}
\end{widetext}

Substituting Eqs. (\ref{eq:S1a-HO}, \ref{eq:S2a-HO}, \ref{eq:DMd-HO}) into Eq. (\ref{eq:CFW2}) and integrating over the initial and the final positions ${(x_i, y_i, x_f, y_f)}$, one can obtain the characteristic function of work
\begin{widetext}
\begin{eqnarray}\label{eq:CFW3-HO}
\chi_W(\nu)&=&\pi\left(\frac{m}{2\pi\hbar}\right)^2
\frac{1}{\sqrt{X_1(\hbar\nu)X(\tau)}}\frac{1}{\sqrt{X(\tau)X_2(\hbar\nu)}}\nonumber \\
&\times& \frac{1}{\sqrt{\frac{m}{2\hbar}\left[\frac{\dot{X}_1(\hbar\nu)}{X_1(\hbar\nu)}+\frac{Y(\tau)}{X(\tau)}\right]}}
\frac{1-e^{-\beta\hbar\omega_0}}{\sqrt{\frac{m}{2\hbar}\left[\frac{\dot{X}(\tau)}{X(\tau)}+\frac{Y_2(\hbar\nu)}{X_2(\hbar\nu)}\right]}}
\sqrt{\frac{m\omega_0}{\hbar\pi\left(1-e^{-2\beta\hbar\omega_0}\right)}}\cdot \Xi,
\end{eqnarray}
where
\begin{eqnarray}\label{eq:CFW4-HO}
\Xi&=&\int dx_i dx_f dy_i\
\exp\Bigg[-\frac{im}{2\hbar}\frac{\left(\frac{x_i}{X_1(\hbar\nu)}+\frac{x_f}{X(\tau)}\right)^2}
{\left(\frac{\dot{X}_1(\hbar\nu)}{X_1(\hbar\nu)}+\frac{Y(\tau)}{X(\tau)}\right)}+
\frac{im}{2\hbar}\frac{\left(\frac{y_i}{X(\tau)}+\frac{x_f}{X_2(\hbar\nu)}\right)^2}
{\left(\frac{\dot{X}(\tau)}{X(\tau)}+\frac{Y_2(\hbar\nu)}{X_2(\hbar\nu)}\right)}
+\frac{im}{2\hbar}\left(\frac{Y_1(\hbar\nu)}{X_1(\hbar\nu)}x^2_i+\frac{\dot{X}(\tau)}{X(\tau)}x^2_f\right) \nonumber \\
&-& \frac{im}{2\hbar}\left(\frac{Y(\tau)}{X(\tau)}y^2_i+\frac{\dot{X}_2(\hbar\nu)}{X_2(\hbar\nu)}x^2_f\right)
-\frac{m\omega_0}{\hbar}\frac{\left(1+e^{-2\beta\hbar\omega_0}\right)
\left(x^2_i+y^2_i\right)-4e^{-\beta\hbar\omega_0}x_i y_i}{2\left(1-e^{-2\beta\hbar\omega_0}\right)}\Bigg].
\end{eqnarray}
\end{widetext}
After some simplification (See Appendix A for details), we obtain the characteristic function of work for the quantum harmonic oscillator with a time-dependent angular frequency
\begin{widetext}
\begin{equation}\label{eq:CFW10-HO}
\chi_W(\nu)=\frac{\sqrt{2}\left(1-e^{-\beta\varepsilon_0}\right)e^{i\nu\Delta\varepsilon/2}}
{\sqrt{Q^*\big(1-e^{2i\nu\varepsilon_1}\big)\big(1-e^{-2\left(i\nu+\beta\right)\varepsilon_0}\big)
+\big(1+e^{2i\nu\varepsilon_1}\big)\big(1+e^{-2\left(i\nu+\beta\right)\varepsilon_0}\big)
-4e^{i\nu\varepsilon_1}e^{-(i\nu+\beta)\varepsilon_0}}},
\end{equation}
where
\begin{equation}\label{eq:CFW9b-HO}
Q^*=\frac{1}{2\omega_0\omega_1}\left\{\omega_0\left[\omega^2_1 X^2(\tau)+\dot{X}^2(\tau)\right]
+\left[\omega^2_1 Y^2(\tau)+\dot{Y}^2(\tau)\right]\right\}.
\end{equation}
\end{widetext}
Please note that Eq. (\ref{eq:CFW10-HO}) is exactly the same as Eq. (17) in Ref. \cite{Deffner2008}. Therefore, the characteristic function of work obtained by the path-integral approach is consistent with the results obtained from solving Schr\"{o}dinger's equation.

\subsection{A free particle inside a rigid box with one wall moving uniformly in time}
As the second model, we consider a rigid box with the left wall fixed at ${x=0}$ while the position of the right wall $x=l_t$ is controlled by an external agent, and moves at a constant velocity ${u}$. Please note that ${l_t}$ plays the role of ${\lambda_t}$ in this model. The forward propagation consists of two parts. For the first part, the right wall of the box does not move and stays at the initial position ${x=l_0}$ during the period of ${t\in[0,\hbar\nu]}$. For the second part, the right wall of the box moves at a constant velocity ${u}$ from the initial position ${l_0}$ at ${t=\hbar\nu}$ to the final position ${l_f=l_0+u\tau}$ at ${t=\tau+\hbar\nu}$. The backward propagation also consists of two parts. While for the first part, the right wall of the box moves at the same velocity ${u}$ from the initial position ${l_0}$ at ${t=0}$ to the final position ${l_f}$ at ${t=\tau}$. For the second part, the right wall of the box does not move and stays at the final position ${l_f}$ during the period of ${t\in[\tau,\tau+\hbar\nu]}$.

In evaluating the propagator (\ref{eq:Forward}) for a given pair of ${(x_i, x_f)}$, there are infinite classical paths (which is quite different from the harmonic oscillator), but we can classify them into four classes \cite{Luz1992}. To be explicit, we can classify the classical paths by specifying which walls (the left or the right) the particle collides with in the first and the last collisions. Class ${\rm I}$: the first collision with the right wall and the last collision with the left wall; Class ${\rm I\!I}$: both the first and the last collisions with the left wall; Class ${\rm I\!I\!I}$: the first collision with the left wall and the last collision with the right wall; Class ${\rm I\!V}$: both the first and the last collisions with the right wall. After some calculations, one can get the actions for the four classes mentioned above \cite{Luz1992}:
\begin{widetext}
\begin{equation}\label{eq:S}
\mathcal{S}^{(n,j)}(x_f,t_f;x_i,t_i)\!=\!\frac{m}{2(t_f-t_i)}\!\left((2nl_0+x+y)^2\!+\!
4nu\left[(xt_f+yt_i)\!+\!nl_0(t_i+t_f) \right]\!+\!4 n^2 u^2 t_i t_f\right), \ \ \ j\!={\rm I,I\!I,I\!I\!I,I\!V}.
\end{equation}
\end{widetext}
Here ${n=1,2,3,...}$ denotes the total number of the collisions of the trajectory, and ${j\!={\rm I,I\!I,I\!I\!I,I\!V}}$ represents the class that the trajectory belongs to. For these four classes, the values of ${(x, y)}$ in Eq. (\ref{eq:S}) are different, and they are given as follows \cite{Luz1992}:
\begin{equation}\label{eq:four-cases1}
({\rm I})\ x=-x_i,\ y=x_f; \ \ ({\rm I\!I})\ x=x_i,\ y=x_f; \nonumber
\end{equation}
\begin{equation}\label{eq:four-cases2}
({\rm I\!I\!I})\ x=x_i,\ y=-x_f; \ \ ({\rm I\!V})\ x=-x_i, y=-x_f. \nonumber
\end{equation}
Then we get the semiclassical propagator by summing up all semiclassical paths of the free particle:
\begin{widetext}
\begin{equation}\label{eq:Gww}
\int^{x_f}_{x_i}\!Dx\ e^{\frac{i}{\hbar}\mathcal{S}[x]}\!=\!
{\left(\frac{m}{2\pi i \hbar (t_f-t_i)}\right)}^{1/2}
\left(\sum^{\infty}_{n=0}e^{\frac{i}{\hbar}\mathcal{S}^{(n,{\rm I})}}
-\sum^{\infty}_{n=0}e^{\frac{i}{\hbar}\mathcal{S}^{(n,{\rm I\!I})}}
+\sum^{\infty}_{n=1}e^{\frac{i}{\hbar}\mathcal{S}^{(n,{\rm I\!I\!I})}}
-\sum^{\infty}_{n=1}e^{\frac{i}{\hbar}\mathcal{S}^{(n,{\rm I\!V})}}\right).
%&=&\frac{2}{\sqrt{l_0 l_f}}e^{\frac{imu}{2\hbar}\left(\frac{x^2_f}{l_f}-\frac{x^2_i}{l_0}\right)}
%\sum^{\infty}_{n=1}e^{\frac{i n^2 \pi^2 \hbar}{2mu}\left(\frac{1}{l_f}-\frac{1}{l_0}\right)}
%\sin\left(\frac{n\pi x_f}{l_f}\right)\sin\left(\frac{n\pi x_i}{l_0}\right).
\end{equation}
\end{widetext}
Please note that the second and the forth class have a minus sign. This is due to the half-wave loss when the particle collides with the walls for odd times \cite{Luz1992}. Also, the forward and the backward propagation can be divided into two independent parts respectively, (see Eqs. (\ref{eq:Forward}-\ref{eq:I3I4})). We will calculate the three parts in Eq. (\ref{eq:CFW2}) first, and then we perform the integration over the initial and the final positions ${(x_i, y_i, x_f, y_f)}$ to calculate the characteristic function of work.

{\it Propagator for the forward propagation.}---For the first part of the forward propagation, ${u=0}$ and the right wall of the box stays at the position of ${x=l_0}$. The particle moves from ${x_i}$ at ${t_i=0}$ to ${x_b}$ at ${t_f=\hbar\nu}$. %From Eq. (\ref{eq:S}) we obtain
%\begin{equation}\label{eq:I1b}
%\mathcal{I}^{(n,j)}_1=\frac{m}{2\hbar\nu}\left(2nl_0+x+y\right)^2,\ \ \ %\ \ \ j=1,2,3,4.
%\end{equation}
Substituting ${u=0}$ and four classes of boundary conditions into Eq. (\ref{eq:Gww}),
\begin{eqnarray}\label{eq:I1c1}
&&({\rm I})\ x\!=\!-x_i,\ y=x_b;\ \ \ \ ({\rm I\!I})\ x=x_i,\ y=x_b;\nonumber \\
&&({\rm I\!I\!I})\ x\!=\!x_i,\ y=-x_b;\ \ ({\rm I\!V})\ x=-x_i,\ y=-x_b;\nonumber
\end{eqnarray}
after some calculations, one can get the analytical expression of the propagator for the first part of the forward propagation:
\begin{widetext}
\begin{equation}\label{eq:I1d}
\int^{x_b}_{x_i} Dx\ e^{\frac{i}{\hbar}I_1[x]}
%{\left(\frac{m}{2\pi i \hbar^2 \nu}\right)}^{1/2}
%\left(\sum^{\infty}_{n=0}e^{\frac{i}{\hbar}\mathcal{I}^{(n,1)}_1}
%-\sum^{\infty}_{n=0}e^{\frac{i}{\hbar}\mathcal{I}^{(n,2)}_1}
%+\sum^{\infty}_{n=1}e^{\frac{i}{\hbar}\mathcal{I}^{(n,3)}_1}
%-\sum^{\infty}_{n=1}e^{\frac{i}{\hbar}\mathcal{I}^{(n,4)}_1}\right)\nonumber \\
=\frac{2}{l_0}\sum^{\infty}_{n=1}
e^{-\frac{i n^2 \pi^2 \hbar^2\nu}{2m l^2_0}}
\sin\left(\frac{n\pi x_i}{l_0}\right)\sin\left(\frac{n\pi x_b}{l_0}\right).
\end{equation}
\end{widetext}
For the second part of the forward propagation, the right wall of the box moves at a constant velocity ${u}$ from the initial position ${l_0}$ to the final position ${l_f}$, the particle moves from ${x_b}$ at ${t_i=0}$ to ${x_f}$ at ${t_f=\tau}$. %Notice that the second part is independent of the first one, the initial time is ${t_i=0}$ but not ${t_i=\hbar\nu}$. Then we obtain
%\begin{equation}\label{eq:I2b}
%\mathcal{I}^{(n,j)}_2=\frac{m}{2\tau}\left[\left(2nl_0+x+y\right)^2+4nu\tau(x+nl_0)\right].
%\end{equation}
Substituting the following four classes of boundary conditions into Eq. (\ref{eq:Gww}),
\begin{eqnarray}\label{eq:I2ca}
&&({\rm I})x=-x_b,\ y=x_f;\ \ \ \ ({\rm I\!I})\ x=x_b,\ y=x_f; \nonumber \\
&&({\rm I\!I\!I})x=x_b,\ y=-x_f;\ \ ({\rm I\!V})\ x=-x_b,\ y=-x_f;\nonumber
\end{eqnarray}
one can get the analytical expression of the propagator for the second part of the forward propagation:
\begin{widetext}
\begin{equation}\label{eq:I2d}
\int^{x_f}_{x_b} Dx\ e^{\frac{i}{\hbar}I_2[x]}
%{\left(\frac{m}{2\pi i \hbar \tau}\right)}^{1/2}
%\left(\sum^{\infty}_{n=0}e^{\frac{i}{\hbar}\mathcal{I}^{(n,1)}_2}
%-\sum^{\infty}_{n=0}e^{\frac{i}{\hbar}\mathcal{I}^{(n,2)}_2}
%+\sum^{\infty}_{n=1}e^{\frac{i}{\hbar}\mathcal{I}^{(n,3)}_2}
%-\sum^{\infty}_{n=1}e^{\frac{i}{\hbar}\mathcal{I}^{(n,4)}_2}\right)\nonumber \\
=\frac{2}{\sqrt{l_0 l_f}}e^{\frac{imu}{2\hbar}\left(\frac{x^2_f}{l_f}-\frac{x^2_b}{l_0}\right)}
\sum^{\infty}_{n=1}e^{\frac{i n^2 \pi^2 \hbar}{2mu}\left(\frac{1}{l_f}-\frac{1}{l_0}\right)}
\sin\left(\frac{n\pi x_b}{l_0}\right)\sin\left(\frac{n\pi x_f}{l_f}\right).
\end{equation}
\end{widetext}
Substituting Eqs. (\ref{eq:I1d}, \ref{eq:I2d}) into Eq. (\ref{eq:Forward}) and integrating over the intermediate position ${x_b}$, one can obtain the result of the propagator (\ref{eq:Forward}) in terms of the imaginary error function Erfi${(x)}$,
\begin{widetext}
\begin{eqnarray}\label{eq:S1a}
\int^{x_f}_{x_i} Dx\ e^{\frac{i}{\hbar}S_1[x]}
&=&\frac{1}{l_0}\sqrt{\frac{\pi\hbar}{2mul_f}}(-1)^{5/4}\nonumber \\
&\times& \sum^{\infty}_{n_1,n_2=1}
e^{-\frac{i n^2_1 \pi^2 \hbar^2\nu}{2m l^2_0}+
\frac{imu x^2_f}{2\hbar l_f}
+\frac{i \pi^2 \hbar}{2mu}\left(\frac{n^2_2}{l_f}+\frac{n^2_1}{l_0}\right)}
\sin\left(\frac{n_1\pi x_i}{l_0}\right)\sin\left(\frac{n_2\pi x_f}{l_f}\right)\cdot A_1(n_1,n_2),
\end{eqnarray}
where
\begin{eqnarray}\label{eq:S1b}
&&A_1(n_1,n_2)=\nonumber \\
&&e^{-\frac{i n_1 n_2 \pi^2 \hbar}{m u l_0}}
\left\{ {\rm E\, r\, f\, i} \left[\left(\frac{1}{2}-\frac{i}{2}\right)\frac{\hbar\pi(n_1-n_2)-mul_0}{\sqrt{m u l_0\hbar}} \right]
-{\rm E\, r\, f\, i} \left[\left(\frac{1}{2}-\frac{i}{2}\right)\frac{\hbar\pi(n_1-n_2)+m u l_0}{\sqrt{m u l_0\hbar}} \right] \right\}\nonumber \\
&&-e^{\frac{i n_1 n_2 \pi^2 \hbar}{m u l_0}}
\left\{{\rm E\, r\, f\, i}
\left[\left(\frac{1}{2}-\frac{i}{2}\right)\frac{\hbar\pi(n_1+n_2)-mul_0}{\sqrt{m u l_0\hbar}} \right]
-{\rm E\, r\, f\, i}
\left[\left(\frac{1}{2}-\frac{i}{2}\right)\frac{\hbar\pi(n_1+n_2)+m u l_0}{\sqrt{m u l_0\hbar}} \right]\right\}.
\end{eqnarray}
\end{widetext}
The definition of the imaginary error function can be found in Ref. \cite{Gradshteyn2007}
\begin{equation}\label{eq:EF}
{\rm E\, r\, f\, i}(x)=-i\cdot {\rm E\, r\, f}
(ix)=-\frac{2i}{\sqrt{\pi}}\int^{ix}_{0}e^{-t^2}dt.
\end{equation}

%\begin{equation}\label{eq:model}
%\int^{x_f}_{x_i} Dx\ e^{\frac{i}{\hbar}S_1[x]}
%=\frac{4}{l_0\sqrt{l_0 l_f}}\sum^{\infty}_{n_1,n_2=1}
%e^{-\frac{i n^2_1 \pi^2 \hbar^2\nu}{2m l^2_0}+
%\frac{imu x^2_f}{2\hbar l_f}
%+\frac{i n^2_2 \pi^2 \hbar}{2mu}\left(\frac{1}{l_f}-\frac{1}{l_0}\right)}
%sin\left(\frac{n_1\pi x_i}{l_0}\right)sin\left(\frac{n_2\pi x_f}{l_f}\right)
%\end{equation}
%\begin{equation}\label{eq:model}
%\times\int^{l_0}_{0} dx_b\ e^{-\frac{imu x^2_b}{2\hbar l_0}}
%sin\left(\frac{n_1\pi x_b}{l_0}\right)sin\left(\frac{n_2\pi x_b}{l_0}\right),
%\end{equation}
%\begin{equation}\label{eq:model}
%\lim_{\nu \to 0,\tau \to 0}\int^{l_0}_{0} dx_i \int^{x_f}_{x_i} Dx\ e^{\frac{i}{\hbar}S_1[x]}=1
%\end{equation}

{\it Propagator for the backward propagation.}---Similarly, one can obtain the expression of the propagator (\ref{eq:Backward}) for the backward propagation,
%For the first part of the backward work process, the piston moves at a %constant velocity ${u}$ from the initial position ${l_0}$ to the final %position ${l_f}$, the particle moves from ${y_0}$ at ${t_i=0}$ to %${y_b}$ at ${t_f=\tau}$, one can obtain
%\begin{widetext}
%\begin{equation}\label{eq:I3b}
%\int^{y_b}_{y_i} Dy\ e^{\frac{i}{\hbar}I_3[y]}
%=\frac{2}{\sqrt{l_0 %l_f}}e^{\frac{imu}{2\hbar}\left(\frac{y^2_b}{l_f}-\frac{y^2_i}{l_0}\right)}
%\sum^{\infty}_{n=1}e^{\frac{i n^2 \pi^2 %\hbar}{2mu}\left(\frac{1}{l_f}-\frac{1}{l_0}\right)}
%sin\left(\frac{n\pi y_i}{l_0}\right)sin\left(\frac{n\pi %y_b}{l_f}\right).
%\end{equation}
%\end{widetext}
%For the second part of the backward work process, ${u=0}$ and the %piston stays in the position of ${y=l_f}$. The particle moves from %${y_b}$ at ${t_i=0}$ to ${y_f}$ at ${t_f=\hbar\nu}$, one can obtain
%\begin{widetext}
%\begin{equation}\label{eq:I4b}
%\int^{y_f}_{y_b} Dy\ e^{\frac{i}{\hbar}I_4[y]}
%=\frac{2}{l_f}\sum^{\infty}_{n=1}
%e^{-\frac{i n^2 \pi^2 \hbar^2\nu}{2m l^2_f}}
%sin\left(\frac{n\pi y_b}{l_f}\right)sin\left(\frac{n\pi %y_f}{l_f}\right).
%\end{equation}
%\end{widetext}
%Bring Eq. (\ref{eq:I3b}, \ref{eq:I4b}) into Eq. (\ref{eq:Backward}) and %integrate over the intermediate position ${y_b}$, one can get
\begin{widetext}
\begin{eqnarray}\label{eq:S2a}
\int^{y_f}_{y_i} Dy\ e^{-\frac{i}{\hbar}S_2[y]}
&=&\frac{1}{l_f}\sqrt{\frac{\pi\hbar}{2mul_0}}(-1)^{5/4}\nonumber \\
&\times& \sum^{\infty}_{n_3,n_4=1}
e^{\frac{i n^2_3 \pi^2 \hbar^2\nu}{2m l^2_f}+
\frac{imu y^2_i}{2\hbar l_0}
+\frac{i \pi^2 \hbar}{2mu}\left(\frac{n^2_3}{l_f}+\frac{n^2_4}{l_0}\right)}
\sin\left(\frac{n_4\pi y_i}{l_0}\right)\sin\left(\frac{n_3\pi y_f}{l_f}\right)\cdot A_2(n_3,n_4),
\end{eqnarray}

where
\begin{eqnarray}\label{eq:S2b}
&&A_2(n_3,n_4)=\nonumber \\
&&e^{-\frac{i n_3 n_4 \pi^2 \hbar}{m u l_f}}
\left\{ {\rm E\, r\, f\, i} \left[\left(\frac{1}{2}-\frac{i}{2}\right)\frac{\hbar\pi(n_3-n_4)-mul_f}{\sqrt{m u l_f\hbar}} \right]
-{\rm E\, r\, f\, i} \left[\left(\frac{1}{2}-\frac{i}{2}\right)\frac{\hbar\pi(n_3-n_4)+m u l_f}{\sqrt{m u l_f\hbar}} \right] \right\}\nonumber \\
&&-e^{\frac{i n_3 n_4 \pi^2 \hbar}{m u l_f}}
\left\{{\rm E\, r\, f\, i}
\left[\left(\frac{1}{2}-\frac{i}{2}\right)\frac{\hbar\pi(n_3+n_4)-mul_f}{\sqrt{m u l_f\hbar}} \right]
-{\rm E\, r\, f\, i}
\left[\left(\frac{1}{2}-\frac{i}{2}\right)\frac{\hbar\pi(n_3+n_4)+m u l_f}{\sqrt{m u l_f\hbar}} \right]\right\}.
\end{eqnarray}
\end{widetext}

%\begin{equation}\label{eq:model}
%\int^{y_f}_{y_i} Dy\ e^{-\frac{i}{\hbar}S_2[y]}
%=\frac{4}{l_f\sqrt{l_0 l_f}}\sum^{\infty}_{n_3,n_4=1}
%e^{\frac{i n^2_3 \pi^2 \hbar^2\nu}{2m l^2_f}+
%\frac{imu y^2_i}{2\hbar l_0}
%+\frac{i n^2_4 \pi^2 \hbar}{2mu}\left(\frac{1}{l_0}-\frac{1}{l_f}\right)}
%sin\left(\frac{n_4\pi y_i}{l_0}\right)sin\left(\frac{n_3\pi x_f}{l_f}\right)
%\end{equation}
%\begin{equation}\label{eq:model}
%\times\int^{l_f}_{0} dy_b\ e^{-\frac{imu y^2_b}{2\hbar l_f}}
%sin\left(\frac{n_3\pi y_b}{l_f}\right)sin\left(\frac{n_4\pi y_b}{l_f}\right)
%\end{equation}

{\it Initial density matrix in the coordinate representation.}---We assume that the system is prepared initially in the thermal equilibrium state,
\begin{equation}\label{eq:DMa}
\hat{\rho}(0)=\sum^{\infty}_{n_5=1} \frac{e^{-\beta E^0_{n_5}}}{Z_0}\left| \Psi_{n_5}(x) \right\rangle \left\langle \Psi_{n_5}(x) \right|,
\end{equation}
where ${Z_0=\sum e^{-\beta E^0_{n_5}}}$ is the partition function. In the coordinate representation,
\begin{eqnarray}\label{eq:DMb}
\rho(x_i,\!y_i)%&=&\sum^{\infty}_{n_5=1}\frac{e^{-\beta E_{n_5}}}{Z}
%\left\langle x_i\left| n_5 \right\rangle \left\langle n_5 \right|y_i\right\rangle
%=\sum^{\infty}_{n_5=1}\frac{e^{-\beta E_{n_5}}}{Z} \Psi_{n_5}(x_i)\Psi^*_{n_5}(y_i)\nonumber \\
\!=\!\frac{2}{Z_0 l_0}\!\sum^{\infty}_{n_5\!=\!1}e^{\!-\!\frac{\beta\!n^2_5\!\pi^2\!\hbar^2}{2ml^2_0}}\!
\!\sin\!\left(\!\frac{n_5\pi x_i}{l_0}\!\right)\!\sin\!\left(\!\frac{n_5\pi y_i}{l_0}\!\right)\!,
\end{eqnarray}
where
\begin{equation}\label{eq:DMc}
E^0_n=\frac{n^2 \pi^2 \hbar^2}{2m l^2_0},\ \ \ \ \ \ \Psi_n(x)=\sqrt{\frac{2}{l_0}}\sin\left(\frac{n\pi x}{l_0}\right)
\end{equation}
are the eigenenergies and the eigenfunctions of the one-dimensional infinite square well.

Substituting Eqs. (\ref{eq:S1a}, \ref{eq:S2a}, \ref{eq:DMb}) into Eq. (\ref{eq:CFW2}) and integrating over the initial and the final positions ${(x_i, y_i, x_f, y_f)}$, one can obtain
\begin{widetext}
\begin{equation}\label{eq:CFW4}
\chi_W(\nu)=\frac{\pi^2\hbar^2}{64m^2u^2l_0l_f Z_0}
\sum^{\infty}_{n_1,n_2,n_3,n_4=1}
e^{-\frac{i n^2_1 \pi^2 \hbar^2\nu}{2m l^2_0}+\frac{i n^2_3 \pi^2 \hbar^2\nu}{2m l^2_f}
-\frac{\beta n^2_1 \pi^2 \hbar^2}{2ml^2_0}}\cdot
A_1(n_1,n_2) A_2(n_3,n_4) A_3(n_4,n_1) A_4(n_2,n_3).
\end{equation}
Here
%\begin{eqnarray}\label{eq:A1}
%A_1(n_1,n_2)=
%e^{-\frac{i n_1 n_2 \pi^2 \hbar}{m u l_0}}
%\left\{ Erfi \left[\left(\frac{1}{2}-\frac{i}{2}\right)\frac{\hbar\pi(n_1-n_2)-mul_0}{\sqrt{m u l_0\hbar}} \right]
%-Erfi \left[\left(\frac{1}{2}-\frac{i}{2}\right)\frac{\hbar\pi(n_1-n_2)+m u l_0}{\sqrt{m u l_0\hbar}} \right] %\right\}\nonumber \\
%-e^{\frac{i n_1 n_2 \pi^2 \hbar}{m u l_0}}
%\left\{Erfi\left[\left(\frac{1}{2}-\frac{i}{2}\right)\frac{\hbar\pi(n_1+n_2)-mul_0}{\sqrt{m u l_0\hbar}} \right]
%-Erfi\left[\left(\frac{1}{2}-\frac{i}{2}\right)\frac{\hbar\pi(n_1+n_2)+m u l_0}{\sqrt{m u l_0\hbar}} \right]\right\},
%\end{eqnarray}
%\begin{eqnarray}\label{eq:A2}
%A_2(n_3,n_4)=
%e^{-\frac{i n_3 n_4 \pi^2 \hbar}{m u l_f}}
%\left\{ Erfi \left[\left(\frac{1}{2}-\frac{i}{2}\right)\frac{\hbar\pi(n_3-n_4)-mul_f}{\sqrt{m u l_f\hbar}} \right]
%-Erfi \left[\left(\frac{1}{2}-\frac{i}{2}\right)\frac{\hbar\pi(n_3-n_4)+m u l_f}{\sqrt{m u l_f\hbar}} \right] %\right\}\nonumber \\
%-e^{\frac{i n_3 n_4 \pi^2 \hbar}{m u l_f}}
%\left\{Erfi\left[\left(\frac{1}{2}-\frac{i}{2}\right)\frac{\hbar\pi(n_3+n_4)-mul_f}{\sqrt{m u l_f\hbar}} \right]
%-Erfi\left[\left(\frac{1}{2}-\frac{i}{2}\right)\frac{\hbar\pi(n_3+n_4)+m u l_f}{\sqrt{m u l_f\hbar}} \right]\right\},
%\end{eqnarray}
\begin{eqnarray}\label{eq:A3}
A_3(n_4,n_1)=e^{\frac{i n_4 n_1 \pi^2 \hbar}{m u l_0}}
\left\{ {\rm E\, r\, f\, i} \left[\left(\frac{1}{2}+\frac{i}{2}\right)\frac{\hbar\pi(n_4-n_1)-mul_0}{\sqrt{m u l_0\hbar}} \right]
-{\rm E\, r\, f\, i} \left[\left(\frac{1}{2}+\frac{i}{2}\right)\frac{\hbar\pi(n_4-n_1)+m u l_0}{\sqrt{m u l_0\hbar}} \right] \right\}\nonumber \\
-e^{-\frac{i n_4 n_1 \pi^2 \hbar}{m u l_0}}
\left\{ {\rm E\, r\, f\, i} \left[\left(\frac{1}{2}+\frac{i}{2}\right)\frac{\hbar\pi(n_4+n_1)-mul_0}{\sqrt{m u l_0\hbar}} \right]
-{\rm E\, r\, f\, i} \left[\left(\frac{1}{2}+\frac{i}{2}\right)\frac{\hbar\pi(n_4+n_1)+m u l_0}{\sqrt{m u l_0\hbar}} \right] \right\},
\end{eqnarray}
\begin{eqnarray}\label{eq:A4}
A_4(n_2,n_3)=e^{\frac{i n_2 n_3 \pi^2 \hbar}{m u l_f}}
\left\{ {\rm E\, r\, f\, i} \left[\left(\frac{1}{2}+\frac{i}{2}\right)\frac{\hbar\pi(n_2-n_3)-mul_f}{\sqrt{m u l_f\hbar}} \right]
-{\rm E\, r\, f\, i} \left[\left(\frac{1}{2}+\frac{i}{2}\right)\frac{\hbar\pi(n_2-n_3)+mul_f}{\sqrt{m u l_f\hbar}} \right] \right\}\nonumber \\
-e^{-\frac{i n_2 n_3 \pi^2 \hbar}{m u l_f}}
\left\{ {\rm E\, r\, f\, i} \left[\left(\frac{1}{2}+\frac{i}{2}\right)\frac{\hbar\pi(n_2+n_3)-mul_f}{\sqrt{m u l_f\hbar}} \right]
-{\rm E\, r\, f\, i} \left[\left(\frac{1}{2}+\frac{i}{2}\right)\frac{\hbar\pi(n_2+n_3)+mul_f}{\sqrt{m u l_f\hbar}} \right] \right\}.
\end{eqnarray}
\end{widetext}
We would like to emphasize that Eq. (\ref{eq:CFW4}) is the analytical expression of the characteristic function of work. This result is obtained based on the path-integral approach. In the following, we will do a self-consistent check to show that Eq. (\ref{eq:CFW4}) is equal to the results obtained based on Schr\"{o}dinger's formalism.

After Fourier transformation, we can get the work distribution for this process:
\begin{widetext}
\begin{eqnarray}\label{eq:PW}
P(W)&=&\frac{1}{2\pi}\int d\nu\ \chi_W(\nu)e^{-i\nu W}=\frac{\pi^2\hbar^2}{64m^2u^2l_0l_f Z_0}\nonumber \\
&\times&\sum^{\infty}_{n_1,n_2,n_3,n_4=1}
\delta\left[W-\frac{\pi^2 \hbar^2}{2m}\left(\frac{n^2_3}{l^2_f}-\frac{n^2_1}{l^2_0}\right)\right]
e^{-\frac{\beta n^2_1 \pi^2 \hbar^2}{2ml^2_0}}\cdot
A_1(n_1,n_2) A_2(n_3,n_4) A_3(n_4,n_1) A_4(n_2,n_3).
\end{eqnarray}
\end{widetext}
%Furthermore, we can check our results from another respect, i.e., the wave function approach.
From Eq. (\ref{eq:PW}), one can see that the physical meaning of ${W}$ in ${P(W)}$ is the ``trajectory work'' associated with the transition process from the initial energy level ${E_{n_1}^0=\frac{n^2_1 \pi^2 \hbar^2}{2m l^2_0}}$ to the final energy level ${E_{n_3}^\tau=\frac{n^2_3 \pi^2 \hbar^2}{2m l^2_f}}$, and the probability of the realization from ${|E_{n_1}^0\rangle}$ to ${|E_{n_3}^\tau\rangle}$ is given by
\begin{widetext}
\begin{equation}\label{eq:TP}
{p(n_1, n_3)\!=\!p_{n_1}\left| \left\langle E^\tau_{n_3} \left| U \right| E^0_{n_1} \right\rangle\right|}^2
\!=\!\frac{\pi^2\hbar^2}{64m^2u^2l_0l_fZ_0}\!\sum^{\infty}_{n_2,n_4=1}e^{-\frac{\beta n^2_1 \pi^2 \hbar^2}{2ml^2_0}}\cdot
A_1(n_1,n_2) A_2(n_3,n_4) A_3(n_4,n_1) A_4(n_2,n_3).
\end{equation}
\end{widetext}
In Appendix B, we prove that the probability of the realization from ${|E_{n_1}^0\rangle}$ to ${|E_{n_3}^\tau\rangle}$ obtained from the path-integral approach (\ref{eq:TP}) is identical to the results obtained by solving the time-dependent Schr\"{o}dinger equation \cite{Quan2011, Doescher1969}.

\section{Path-integral approach to the calculation of the work statistics in the classical systems}
From these two models, one can see that one important feature in the calculation of the characteristic function of work is that there are two propagations in quantum regime, i.e., the forward propagation and the backward propagation, while there's only one propagation in classical regime. Notice that when $\hbar\to 0$, the forward and the backward propagation will converge, and one can prove that they converge to the classical trajectory. We can rewrite Eq. (\ref{eq:CFW2}) into the following form \cite{Ken2018}
\begin{equation}\label{eq:CFW2A}
\chi_W(\nu)\!=\!\int\!e^{\frac{i}{\hbar}\left(S_2[x]-S_2[y]\right)} \rho(x_i,y_i)e^{i\nu W_{\nu}[x]},
\end{equation}
where
\begin{equation}\label{eq:CFW2B}
W_{\nu}[x]=\int^{\tau}_{0}dt\frac{1}{\hbar\nu}\int^{\hbar\nu}_{0}ds\dot{\lambda}_t\frac{\partial V[\lambda_t,x(t+s)]}{\partial\lambda_t}
\end{equation}
is the quantum work functional, and ${e^{\frac{i}{\hbar}\left(S_2[x]-S_2[y]\right)} \rho(x_i,y_i)}$ is the quasi-probability associated with this work functional. We define ${\zeta=(x+y)/2}$ and ${\gamma=x-y}$. After doing the stationary phase approximation, the characteristic function of work (\ref{eq:CFW2A}) becomes \cite{Ken2018}
\begin{widetext}
\begin{equation}\label{eq:CFW2C}
\chi^{cl}_W(\nu)=\int d\zeta_f d\zeta_i \int DX \delta(m\ddot{\zeta}(t)+V'[\zeta(t)]) e^{i\nu W_{cl}[X]}P(\zeta_i,\dot{\zeta}_i),
\end{equation}
\end{widetext}
where
\begin{equation}\label{eq:CFW2D}
W_{cl}[\zeta]=\int^{\tau}_{0}dt\dot{\lambda}_t\frac{\partial V[\lambda_t,\zeta(t)]}{\partial\lambda_t}
\end{equation}
is the classical work functional \cite{Sekimoto2010}, and
\begin{equation}\label{eq:CFW2E}
P(\zeta_i,\dot{\zeta}_i)=\frac{1}{\pi\hbar}\int d\gamma_i e^{-(i/\hbar)M\gamma_i\dot{\zeta}_i}\rho(\zeta_i,\gamma_i)
\end{equation}
is the classical probability distribution of the initial state. Comparing Eq. (\ref{eq:CFW2C}) with Eq. (\ref{eq:CFW2A}), one can see that when we take the classical limit, the two quantum paths ${x(t)}$ and ${y(t)}$ converge to the classical trajectory ${\zeta(t)}$ which satisfies Newton's equation ${m\ddot{\zeta}(t)}$ ${+V'[\zeta(t)]=0}$. The quantum work functional ${W_{\nu}[x]}$ converges to the classical work functional ${W_{cl}[\zeta]}$, and the initial density matrix ${\rho(x_i,y_i)}$ converges to the classical probability distribution ${P(\zeta_i,\dot{\zeta}_i)}$ in the phase space. The integral over the initial and the final positions is replaced by the integral over the generalized coordinates in the phase space.

Next, we will calculate the classical work distribution in the model of a harmonic oscillator with a time-dependent angular frequency and a free particle inside an expanding piston.

\subsection{Classical harmonic oscillator with a time-dependent angular frequency}
As for the model of a classical harmonic oscillator with a time-dependent angular frequency, one can calculate the classical work functional through its definition Eq. (\ref{eq:CFW2D}),
\begin{equation}\label{eq:CFW23}
W_{cl}\!=\!\int^{\tau}_{0}dt\ m\omega(t)\dot{\omega}(t)\!\left[\dot{x}_i X\!(t)\!+\!x_i Y\!(t)\right]^2,
\end{equation}
where ${x_i}$ and ${\dot{x}_i}$ are the initial position and the initial velocity of the harmonic oscillator, ${X(t)}$ and ${Y(t)}$ are two linearly independent solutions of the classical equation of motion ${\ddot{x}(t)}$${=}$ ${-\omega^2(t)x(t)}$. We assume that the system is in a thermal equilibrium state initially, so the classical probability distribution is given by
\begin{equation}\label{eq:CFW24}
P(x_i,\dot{x}_i)=\frac{\beta m\omega_0}{2\pi}e^{-\frac{\beta m}{2}(\dot{x}_i^2+\omega_i^2 x_i^2)}.
\end{equation}
Thus the classical characteristic function of work given by Eq. (\ref{eq:CFW2C}) can be written as
\begin{eqnarray}\label{eq:CFW25}
\chi^{cl}_W(\nu)&\!=\!&\frac{\beta m\omega_0}{2\pi}\!\int^{\infty}_{-\infty}dx_i\!\int^{\infty}_{-\infty}d\dot{x}_i \
e^{-\frac{\beta m}{2}(\dot{x}_i^2+\omega_0^2 x_i^2)}\ \\ \nonumber
&&\times e^{i\nu\int^{\tau}_{0}dt \ m \ \omega(t)\dot{\omega}(t)\left[\dot{x_i} X(t)+x_i Y(t)\right]^2}.
\end{eqnarray}
After some calculations one can obtain
\begin{equation}\label{eq:CFW26}
\chi^{cl}_W(\nu)=\frac{\beta\omega_0}{\sqrt{A\nu^2+B\nu+C}},
\end{equation}
where ${A\!=\!2Q^*\omega_0\omega_1-\omega_0^2-\omega_1^2}$, ${B\!=\!2i\left(\beta\omega_0^2-Q^*\beta\omega_0\omega_1\right)}$, ${C=\beta^2\omega_0^2}$. This result is identical to Eq. (25) in Ref. \cite{Deffner2008}. We would like to emphasize that we obtain the classical characteristic function of work (Eq. (\ref{eq:CFW26})) by doing path integral in the classical trajectory space. But in Ref. \cite{Deffner2008}, this result was obtained by taking the classical limit (${\hbar\to 0}$) of the quantum characteristic function of work (\ref{eq:CFW10-HO}). We believe that in comparison with the method in Ref. \cite{Deffner2008}, our method encodes more information about the quantum-to-classical transition at the level of individual trajectories.

\begin{figure*}[htbp]
\begin{center}
\includegraphics[width=0.75\textwidth]{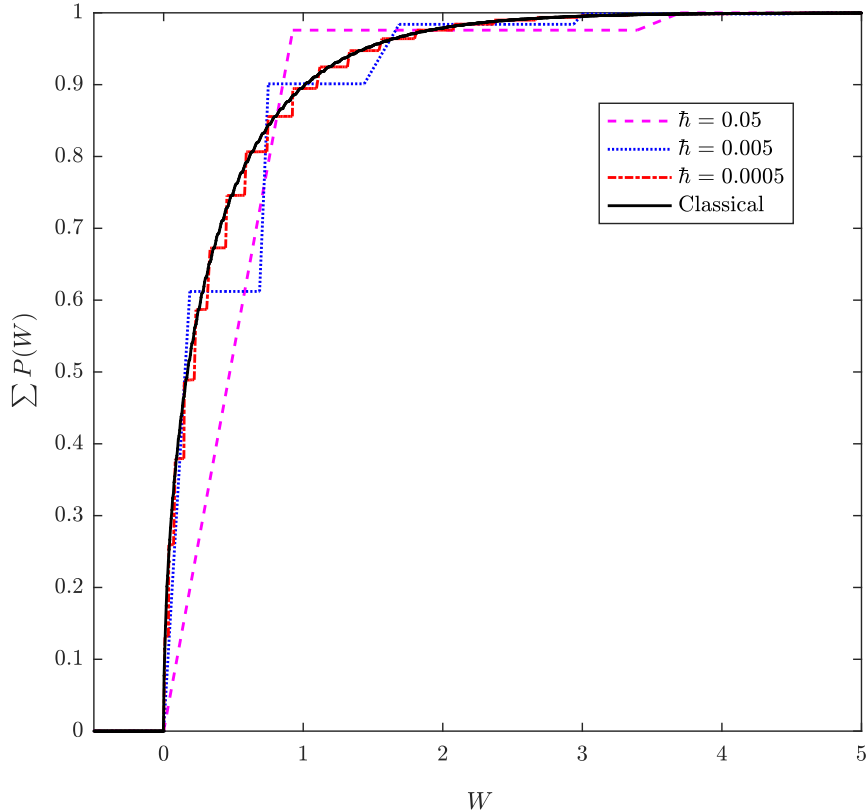}
\caption{(Color online).
Quantum-to-classical transition of the accumulated work distribution of the expanding piston model. Here the parameters are chosen as: ${m=1}$, ${l_0=0.01}$, ${l_f=0.02}$, ${u=0.01}$. The black curve corresponds to the classical accumulated work distribution which is defined as ${\int_{-\infty}^{W} dW' P(W')}$, where ${P(W)}$ is given by Eq. (13) of Ref. \cite{Lua2005}. Other curves correspond to quantum accumulated work distribution defined as ${\sum^{W'=W}_{W'=\{W\}_{min}}P(W')}$, where ${P(W)}$ is given by Eq. (\ref{eq:PW}). The initial states for both quantum and classical cases are thermal equilibrium states at ${\beta=1}$, and the values of ${\hbar}$ are chosen to be ${\hbar=0.05, 0.005, 0.0005}$.} \label{QCC}
\end{center}
\end{figure*}

\subsection{A free particle inside a rigid box with one wall moving uniformly in time}
For the model of a free particle inside a rigid box with the right wall  moving uniformly in time, the usual definition of the trajectory work (\ref{eq:CFW2D}) is not applicable \cite{Gong2016}. The classical work functional is given by Eq. (10) in Ref. \cite{Lua2005}, and the work distribution can be calculated by employing the path-integral approach (see Eq. (13) in Ref. \cite{Lua2005}). Due to this fact, the above analysis (Eqs. (\ref{eq:CFW2A}- \ref{eq:CFW2E})) is not applicable to the piston system. In order to demonstrate the quantum classical correspondence of non-equilibrium work, we have to find other evidences. For example, if Eq. (\ref{eq:PW}) converges to Eq. (13) of Ref. \cite{Lua2005} in the classical limit ${\hbar\to 0}$, we can demonstrate the quantum-to-classical transition at the level of individual trajectories. Considering the fact that it is difficult to get the analytical expression of Eq. (\ref{eq:PW}) when taking the limit ${\hbar\to 0}$, we refer to numerical simulations. Due to the discreteness of the distribution of quantum work, we calculate the accumulated work distribution instead of the work distribution itself \cite{Gong2014}. In Fig. \ref{QCC}, we plot the classical accumulated work distribution (black curve), as well as the quantum accumulated work distribution from Eq. (\ref{eq:PW}) for various values of ${\hbar}$. One can see that there are many stairs in the curves for quantum accumulated work distribution due to the discreteness, but when we decrease the value of ${\hbar}$, the quantum accumulated work distribution will become closer and closer to its classical counterpart. It is obvious that the quantum accumulated work distribution will converge to the classical accumulated work distribution in the classical limit ${\hbar\to 0}$. Thus we demonstrate numerically the quantum-to-classical transition of the work statistics in the expanding piston model.

\section{Discussion and Conclusion}
Feynman's path integral formalism provides important insights to the understanding of quantum mechanics. Usually the trajectory work in an isolated quantum system is defined via the so-called two-point measurement. While in classical systems, the trajectory work is defined via trajectories in the phase space. In this work, we bridge the two regimes by utilizing the path-integral approach. To be more explicit, we study the calculation of work statistics by utilizing the path-integral approach in both quantum and classical systems. We have obtained the analytical work distributions for two prototype quantum systems by the path-integral approach, and we prove that they are equivalent to the Schr\"{o}dinger's formalism. We consider a quantum harmonic oscillator with a time-dependent angular frequency, and a free particle inside a rigid box with one wall moving uniformly in time. For these two systems, the semiclassical propagators are exact, i.e., only classical paths contribute to the propagator and the calculation is significantly simplified. We sum over all the possible classical paths and get the characteristic function of work. Even though the evaluation of path integral in many cases are cumbersome tasks, it is straightforward to extend the method to open quantum systems \cite{Ken2018} and quantum fields.

In addition, we also calculate the work statistics in classical systems by utilizing the path-integral approach. Thus the path-integral approach can be regarded as a unified method, in principle, for the calculation of the work statistics in both quantum and classical systems. Our results provide good examples to show the effectiveness of the path-integral approach to the calculation of work statistics in both quantum and classical systems, and may shed new light on the physical meaning of quantum trajectory work. Furthermore, we show the quantum-to-classical transition at the level of individual trajectory.

Many questions remain open in the field of path-integral approach to quantum stochastic thermodynamics, such as the path-integral approach to quantum fluctuation theorem in quantum fields \cite{Bartolotta2018, Ortega2019, Dong2019}. Further studies in this line will be given in forth coming papers.

\section*{Acknowledgment}
H. T. Quan acknowledges support from the National Science Foundation of China under grants 11775001, 11534002, and 11825001.

\appendix
\section{Details for the derivation of Eq. (\ref{eq:CFW10-HO}) from Eqs. (\ref{eq:CFW3-HO}, \ref{eq:CFW4-HO})}
The integrations in Eq. (\ref{eq:CFW4-HO}) can be calculated analytically and we obtain
\begin{widetext}
\begin{equation}\label{eq:CFW5-HO}
\Xi=\sqrt{\frac{\pi}{2}}(1-i)\frac{1}{\sqrt{\frac{m}{2\hbar}C}}
\frac{\sqrt{\pi}}{\sqrt{C_1+iC_2-iC_3-iC_4}}\frac{\sqrt{\pi}}{\sqrt{C_1-C_5+iC_6-iC_7-iC_8}},
\end{equation}
where
\begin{subequations}\label{eq:CFW5a-HO}
\begin{equation}
C=\frac{1}{X^2(\tau)A}-\frac{1}{X^2_2(\hbar\nu)B}+\frac{\dot{X}_2(\hbar\nu)}{X_2(\hbar\nu)}-\frac{\dot{X}(\tau)}{X(\tau)},\nonumber
\end{equation}
\begin{equation}
C_1=\frac{m\omega_0}{2\hbar}\frac{1+e^{-2\beta\hbar\omega_0}}{1-e^{-2\beta\hbar\omega_0}},\ \ \
C_2=\frac{m}{2\hbar}\frac{1}{X^2_1(\hbar\nu)A},\ \ \ C_3=\frac{m}{2\hbar}\frac{Y_1(\hbar\nu)}{X_1(\hbar\nu)},\ \ \
C_4=\frac{m}{2\hbar}\frac{1}{C X^2_1(\hbar\nu)X^2(\tau)A^2},\nonumber
\end{equation}
\begin{equation}
C_5=\frac{\left(\frac{2m\omega_0}{\hbar}\frac{e^{-\beta\hbar\omega_0}}{1-e^{-2\beta\hbar\omega_0}}
-\frac{im}{\hbar}\frac{1}{C X_1(\hbar\nu)X(\tau)X(\tau)X_2(\hbar\nu)AB}\right)^2}
{4\left(\frac{m\omega_0}{\hbar}\frac{1+e^{-2\beta\hbar\omega_0}}{1-e^{-2\beta\hbar\omega_0}}
+\frac{im}{2\hbar}\frac{1}{X^2_1(\hbar\nu)A}-\frac{im}{2\hbar}\frac{Y_1(\hbar\nu)}{X_1(\hbar\nu)}
-\frac{im}{2\hbar}\frac{1}{C X^2_1(\hbar\nu)X^2(\tau)A^2}\right)},\nonumber
\end{equation}
\begin{equation}
C_6=\frac{m}{2\hbar}\frac{Y(\tau)}{X(\tau)},\ \ \
C_7=\frac{m}{2\hbar}\frac{1}{X^2(\tau)B},\ \ \
C_8=\frac{m}{2\hbar}\frac{1}{C X^2(\tau)X^2_2(\hbar\nu)B^2},\nonumber
\end{equation}
\end{subequations}
and
\begin{equation}\label{eq:CFW5b-HO}
A=\frac{\dot{X}_1(\hbar\nu)}{X_1(\hbar\nu)}+\frac{Y(\tau)}{X(\tau)},\ \ \ \ \ \ \
B=\frac{\dot{X}(\tau)}{X(\tau)}+\frac{Y_2(\hbar\nu)}{X_2(\hbar\nu)}.\nonumber
\end{equation}
Substituting these results into Eq. (\ref{eq:CFW3-HO}), we get
\begin{equation}\label{eq:CFW6-HO}
\chi_W(\nu)=\sqrt{2}\left(1-e^{-\beta\hbar\omega_0}\right)/\sqrt{D},
\end{equation}
where
\begin{eqnarray}\label{eq:CFW7a-HO}
D&=&\left(1+e^{-2\beta\hbar\omega_0}\right)\Bigg[\Bigg(
\frac{\dot{X}_1(\hbar\nu)Y_1(\hbar\nu)X(\tau)\dot{X}(\tau)X_2(\hbar\nu)}{X_1(\hbar\nu)}
+\frac{\dot{X}_1(\hbar\nu)Y_1(\hbar\nu)X^2(\tau)Y_2(\hbar\nu)}{X_1(\hbar\nu)} \nonumber \\
&+&Y_1(\hbar\nu)\dot{X}(\tau)Y(\tau)X_2(\hbar\nu)+Y_1(\hbar\nu)X(\tau)Y(\tau)Y_2(\hbar\nu)
-\dot{X}_1(\hbar\nu)\dot{X}(\tau)Y(\tau)X_2(\hbar\nu)-\dot{X}_1(\hbar\nu)X(\tau)Y(\tau)Y_2(\hbar\nu)\nonumber \\
&-&\frac{X_1(\hbar\nu)\dot{X}(\tau)Y^2(\tau)X_2(\hbar\nu)}{X(\tau)}
-X_1(\hbar\nu)Y^2(\tau)Y_2(\hbar\nu)+\dot{X}_1(\hbar\nu)X_2(\hbar\nu)
+\frac{X_1(\hbar\nu)Y(\tau)X_2(\hbar\nu)}{X(\tau)} \nonumber \\
&-&\frac{X(\tau)\dot{X}(\tau)X_2(\hbar\nu)}{X_1(\hbar\nu)}
-\frac{X^2(\tau)Y_2(\hbar\nu)}{X_1(\hbar\nu)}\Bigg)C
+\frac{\dot{X}_1(\hbar\nu)X(\tau)+X_1(\hbar\nu)Y(\tau)}{X_2(\hbar\nu)\dot{X}(\tau)+Y_2(\hbar\nu)X(\tau)}
+\frac{X_2(\hbar\nu)\dot{X}(\tau)+Y_2(\hbar\nu)X(\tau)}{\dot{X}_1(\hbar\nu)X(\tau)+X_1(\hbar\nu)Y(\tau)}\Bigg]
\nonumber \\
&+&i\omega_0 \left(1-e^{-2\beta\hbar\omega_0}\right)
\Big[\dot{X}_1(\hbar\nu)X(\tau)\dot{X}(\tau)X_2(\hbar\nu)
+\dot{X}_1(\hbar\nu)X^2(\tau)Y_2(\hbar\nu) \nonumber \\
&+&X_1(\hbar\nu)\dot{X}(\tau)Y(\tau)X_2(\hbar\nu)
+X_1(\hbar\nu)X(\tau)Y(\tau)Y_2(\hbar\nu)\Big]C \nonumber \\
&+&\frac{i}{\omega_0}\left(1-e^{-2\beta\hbar\omega_0}\right)\Bigg[
\frac{\dot{X}_1(\hbar\nu)Y_1(\hbar\nu)\dot{X}(\tau)Y(\tau)X_2(\hbar\nu)}{X_1(\hbar\nu)}
+\frac{\dot{X}_1(\hbar\nu)Y_1(\hbar\nu)X(\tau)Y(\tau)Y_2(\hbar\nu)}{X_1(\hbar\nu)} \nonumber \\
&+&\frac{Y_1(\hbar\nu)\dot{X}(\tau)Y^2(\tau)X_2(\hbar\nu)}{X(\tau)}
+Y_1(\hbar\nu)Y^2(\tau)Y_2(\hbar\nu)
-\frac{\dot{X}(\tau)Y(\tau)X_2(\hbar\nu)}{X_1(\hbar\nu)}
-\frac{X(\tau)Y(\tau)Y_2(\hbar\nu)}{X_1(\hbar\nu)}\nonumber \\
&-&\frac{\dot{X}_1(\hbar\nu)Y_1(\hbar\nu)X_2(\hbar\nu)}{X_1(\hbar\nu)}
-\frac{Y_1(\hbar\nu)Y(\tau)X_2(\hbar\nu)}{X(\tau)}
+\frac{X_2(\hbar\nu)}{X_1(\hbar\nu)}\Bigg]C \nonumber \\
&+&\frac{i}{\omega_0}\left(1-e^{-2\beta\hbar\omega_0}\right)\Bigg[
\frac{\dot{X}(\tau)Y(\tau)X_2(\hbar\nu)+X(\tau)Y(\tau)Y_2(\hbar\nu)}{\dot{X}_1(\hbar\nu)X^2(\tau)+X_1(\hbar\nu)X(\tau)Y(\tau)}
-\frac{X_2(\hbar\nu)}{\dot{X}_1(\hbar\nu)X^2(\tau)+X_1(\hbar\nu)X(\tau)Y(\tau)} \nonumber \\
&-&\frac{\dot{X}_1(\hbar\nu)Y_1(\hbar\nu)X(\tau)+X_1(\hbar\nu)Y_1(\hbar\nu)Y(\tau)}{X_1(\hbar\nu)\dot{X}(\tau)X_2(\hbar\nu)+X_1(\hbar\nu)X(\tau)Y_2(\hbar\nu)}
-\frac{X(\tau)}{X_1(\hbar\nu)\dot{X}(\tau)X_2(\hbar\nu)+X_1(\hbar\nu)X(\tau)Y_2(\hbar\nu)}\Bigg]-4e^{-\beta\hbar\omega_0},
\end{eqnarray}
and
\begin{equation}\label{eq:CFW7b-HO}
C=\frac{X_1(\hbar\nu)}{\dot{X}_1(\hbar\nu)X^2(\tau)+X_1(\hbar\nu)X(\tau)Y(\tau)}
-\frac{X(\tau)}{\dot{X}(\tau)X^2_2(\hbar\nu)+X(\tau)X_2(\hbar\nu)Y_2(\hbar\nu)}
+\frac{\dot{X}_2(\hbar\nu)}{X_2(\hbar\nu)}-\frac{\dot{X}(\tau)}{X(\tau)}.
\end{equation}
Substituting the expressions of ${X_1(\hbar\nu)}$, ${Y_1(\hbar\nu)}$ (\ref{eq:XY-HO2}), ${X_2(\hbar\nu)}$ and ${Y_2(\hbar\nu)}$ (\ref{eq:XY2-HO2}) into Eq. (\ref{eq:CFW7a-HO}) and Eq. (\ref{eq:CFW7b-HO}), one can obtain
\begin{eqnarray}\label{eq:CFW8a-HO}
D&=&\left(1+e^{-2\beta\hbar\omega_0}\right)\Bigg[
\frac{1}{\omega_1}\cos{\nu\varepsilon_0}\sin{\nu\varepsilon_1}
-\frac{1}{\omega_0\omega_1}\sin{\nu\varepsilon_0}\sin{\nu\varepsilon_1}Y(\tau)\dot{Y}(\tau)
-\frac{\omega_0\sin{\nu\varepsilon_1}}{\omega_1\sin{\nu\varepsilon_0}}X(\tau)\dot{X}(\tau)\nonumber \\
&-&\frac{\omega_0\cos{\nu\varepsilon_1}}{\sin{\nu\varepsilon_0}}X^2(\tau)
+\frac{\omega_0\cos^2{\nu\varepsilon_0}\sin{\nu\varepsilon_1}}{\omega_1\sin{\nu\varepsilon_0}}X(\tau)\dot{X}(\tau)
+\frac{\omega_0\cos^2{\nu\varepsilon_0}\cos{\nu\varepsilon_1}}{\sin{\nu\varepsilon_0}}X^2(\tau)
-\frac{1}{\omega_0}\sin{\nu\varepsilon_0}\cos{\nu\varepsilon_1}Y^2(\tau)\Bigg]C\nonumber \\
&+&\left(1+e^{-2\beta\hbar\omega_0}\right)\Bigg[
\frac{\omega_1\cos{\nu\varepsilon_0}X(\tau)+\omega_1/\omega_0\sin{\nu\varepsilon_0}Y(\tau)}
{\omega_1\cos{\nu\varepsilon_1}X(\tau)+\sin{\nu\varepsilon_1}\dot{X}(\tau)}
+\frac{\omega_0\cos{\nu\varepsilon_1}X(\tau)+\omega_0/\omega_1\sin{\nu\varepsilon_1}\dot{X}(\tau)}
{\omega_0\cos{\nu\varepsilon_0}X(\tau)+\sin{\nu\varepsilon_0}Y(\tau)}\Bigg]\nonumber \\
&+&\left(1-e^{-2\beta\hbar\omega_0}\right)\Bigg[
\frac{i\omega_0}{\omega_1}\cos{\nu\varepsilon_0}\sin{\nu\varepsilon_1}X(\tau)\dot{X}(\tau)
+i\omega_0\cos{\nu\varepsilon_0}\cos{\nu\varepsilon_1}X^2(\tau)\nonumber \\
&+&\frac{i}{\omega_0\omega_1}\cos{\nu\varepsilon_0}\sin{\nu\varepsilon_1}Y(\tau)\dot{Y}(\tau)
+\frac{i}{\omega_0}\cos{\nu\varepsilon_0}\cos{\nu\varepsilon_1}Y^2(\tau)
-\frac{i\cos^2{\nu\varepsilon_0}\sin{\nu\varepsilon_1}}{\omega_1\sin{\nu\varepsilon_0}}
+\frac{i\sin{\nu\varepsilon_1}}{\omega_1\sin{\nu\varepsilon_0}}\Bigg]C-4e^{-\beta\hbar\omega_0} \nonumber \\
&+&\left(1-e^{-2\beta\hbar\omega_0}\right)\Bigg[
\frac{i\cos{\nu\varepsilon_1}Y(\tau)+1/\omega_1\sin{\nu\varepsilon_1}\dot{Y}(\tau)}
{\omega_0\cos{\nu\varepsilon_0}X(\tau)+\sin{\nu\varepsilon_0}Y(\tau)}
+\frac{i\omega_1\sin{\nu\varepsilon_0}X(\tau)-i\omega_1/\omega_0\cos{\nu\varepsilon_0}Y(\tau)}
{\sin{\nu\varepsilon_1}\dot{X}(\tau)+\omega_1\cos{\nu\varepsilon_1}X(\tau)}\Bigg],
\end{eqnarray}
where
\begin{equation}\label{eq:CFW8b-HO}
C=\frac{\omega_1\cos{\nu\varepsilon_1}\dot{X}(\tau)-\omega^2_1\sin{\nu\varepsilon_1}X(\tau)}
{\sin{\nu\varepsilon_1}\dot{X}(\tau)+\omega_1\cos{\nu\varepsilon_1}X(\tau)}
-\frac{\omega_0\cos{\nu\varepsilon_0}\dot{X}(\tau)+\sin{\nu\varepsilon_0}\dot{Y}(\tau)}
{\omega_0\cos{\nu\varepsilon_0}X(\tau)+\sin{\nu\varepsilon_0}Y(\tau)}.
\end{equation}
After further simplifications, we obtain
\begin{equation}\label{eq:CFW9-HO}
D=Q^*\big(1-e^{2i\nu\varepsilon_1}\big)\big(1-e^{-2\left(i\nu+\beta\right)\varepsilon_0}\big)e^{-i\nu\Delta\varepsilon}
+\big(1+e^{2i\nu\varepsilon_1}\big)\big(1+e^{-2\left(i\nu+\beta\right)\varepsilon_0}\big)e^{-i\nu\Delta\varepsilon}
-4e^{-\beta\hbar\omega_0},
\end{equation}
where
\begin{equation}\label{eq:CFW9c-HO}
Q^*=\frac{1}{2\omega_0\omega_1}\left\{\omega_0\left[\omega^2_1 X^2(\tau)+\dot{X}^2(\tau)\right]
+\left[\omega^2_1 Y^2(\tau)+\dot{Y}^2(\tau)\right]\right\}.
\end{equation}
\end{widetext}
Finally, by substituting Eq. (\ref{eq:CFW9-HO}) into Eq. (\ref{eq:CFW6-HO}) we obtain Eq. (\ref{eq:CFW10-HO}).

\section{Alternative derivation of Eq.(\ref{eq:TP}) from solving Schr\"{o}dinger equation}
There are exact solutions to the time-dependent Schr\"{o}dinger equation for the expanding piston model \cite{Quan2011, Doescher1969}. The time-dependent Schr\"{o}dinger equation is
\begin{equation}\label{eq:SE}
H(t)\Psi=i\hbar\frac{\partial \Psi}{\partial t},
\end{equation}
where the Hamiltonian characterizes the rigid box with a moving wall. The exact solutions to Eq. (\ref{eq:SE}) read
\begin{equation}\label{eq:solution1}
\phi_n(x,t)\!=\!\sqrt{\frac{2}{l(t)}}
e^{i\alpha\xi\left(\frac{x}{l(t)}\right)^2\!-\!\frac{i n^2 \pi^2}{4\alpha}\!\left(\!1-\frac{1}{\xi}\!\right)}
\sin\!\left(\!\frac{n\pi x}{l(t)}\!\right),
\end{equation}
where ${\xi=l(t)/l_0}$ and ${\alpha=mul_0/(2\hbar)}$. These functions vanishing at ${x=0}$ and ${x=l(t)}$ as required, remain normalized as the right wall at $x=l(t)$ moves, and form a complete orthogonal set. Thus any wave function can be expanded in terms of them,
\begin{equation}\label{eq:expand}
\Psi(x,t)=\sum_{n}a_n\phi_n(x,t).
\end{equation}
The expansion coefficients ${a_n}$ remains constant as the right wall moves, with their values being determined by the wave function at ${t=0}$ in the usual manner,
\begin{equation}\label{eq:an}
a_n=\int^{l_0}_{0}\phi^*_n(x,t)\Psi(x,0)dx.
\end{equation}
The system stays in ${E_{n_1}^0}$ at the beginning, so the initial state reads
\begin{equation}\label{eq:Psi0}
\Psi(x,0)=\sqrt{\frac{2}{l_0}} \sin\left(\frac{n_1\pi x}{l_0}\right),
\end{equation}
then we have
\begin{equation}\label{eq:an1}
a_n=\frac{2}{l_0}\int^{l_0}_{0}
e^{-\frac{i mux^2}{2\hbar l_0}}
\sin\!\left(\!\frac{n\pi x}{l_0}\!\right)\!\sin\!\left(\!\frac{n_1\pi x}{l_0}\!\right)dx.
\end{equation}
In order to find the transition probability from ${E_{n_1}^0}$ to ${E_{n_3}^\tau}$, one must re-expand the wave function in terms of the instantaneous energy eigenfunctions ${u_k(x,t)}$,
\begin{equation}\label{eq:re-expand}
\Psi(x,t)=\sum_{n}a_n\phi_n(x,t)=\sum_{k}C_k(t) u_k(x,t),
\end{equation}
where
\begin{equation}\label{eq:uk}
u_k(x,t)=\sqrt{\frac{2}{l(t)}} \sin\left(\frac{k\pi x}{l(t)}\right).
\end{equation}
Then we have
\begin{equation}\label{eq:Ck}
C_k(t)=\sum_{n}a_n \int^{l(t)}_{0}u_k(x,t)\phi_n(x,t)dx.
\end{equation}
When ${k=n_3}$, one can obtain the transition probability amplitude from ${E_{n_1}^0}$ to ${E_{n_3}^\tau}$ by substituting Eqs. (\ref{eq:solution1}, \ref{eq:an1}, \ref{eq:uk}) into Eq. (\ref{eq:Ck}),
\begin{equation}\label{eq:Cn3}
C_{n_3}(t_f)=\frac{4}{l_0 l_f}\sum_{n_2} J_1 J_2 \cdot
e^{-\frac{i n^2 \pi^2 \hbar}{2mu}\left(\frac{1}{l_0}-\frac{1}{l_f}\right)},
\end{equation}
where
\begin{equation}\label{eq:PTI1}
J_1=\int^{l_0}_{0} e^{-\frac{i mux^2}{2\hbar l_0}}
\sin\left(\frac{n_1\pi x}{l_0}\right)\sin\left(\frac{n_2\pi x}{l_0}\right)dx, \nonumber
\end{equation}
\begin{equation}\label{eq:PTI1a}
J_2=\int^{l_f}_{0} e^{\frac{i mux^2}{2\hbar l_f}}
\sin\left(\frac{n_2\pi x}{l_0}\right)\sin\left(\frac{n_3\pi x}{l_0}\right)dx. \nonumber
\end{equation}
Thus the transition probability is given by
\begin{equation}\label{eq:TP1}
\left|C_{n_3}(t_f)\right|^2\!=\!\frac{16}{l^2_0 l^2_f}\!\sum_{n_2,n_4} J_1 J_2 J_3 J_4\!\cdot
e^{\frac{i \pi^2 \hbar}{2mu}\!\left(\!\frac{1}{l_0}-\frac{1}{l_f}\!\right)\!\left(\!n^2_4-n^2_2\!\right)},
\end{equation}
where
\begin{eqnarray}\label{eq:PTI2}
J_3&=&\int^{l_0}_{0} e^{\frac{i mux^2}{2\hbar l_0}}
\sin\left(\frac{n_4\pi x}{l_0}\right)\sin\left(\frac{n_1\pi x}{l_0}\right)dx, \nonumber \\
J_4&=&\int^{l_f}_{0} e^{-\frac{i mux^2}{2\hbar l_f}}
\sin\left(\frac{n_3\pi x}{l_0}\right)\sin\left(\frac{n_4\pi x}{l_0}\right)dx. \nonumber
\end{eqnarray}
After performing these integrations, one can obtain
\begin{equation}\label{eq:TP2}
\left|C_{n_3}(t_f)\right|^2=
\frac{\pi^2\hbar^2}{64m^2u^2l_0l_f}\sum^{\infty}_{n_2,n_4=1}
A_1 A_2 A_3 A_4.
\end{equation}
Here ${A_i}$, ${i=1,2,3,4}$ is defined in Eqs. (\ref{eq:S1b}, \ref{eq:S2b}, \ref{eq:A3}, \ref{eq:A4}). Furthermore, we know that the system is prepared in a thermal equilibrium state initially, so the probability of finding the system in ${E_{n_1}^0}$ is ${e^{-\beta E_{n_1}^0}/Z_0}$, and thus the probability of the realization from ${|E_{n_1}^0\rangle}$ to ${|E_{n_3}^\tau\rangle}$ is given by Eq. (\ref{eq:TP}). Thus we conclude that the work distribution obtained by the path integral approach is consistent with the results obtained from solving Schr\"{o}dinger's equation.

\end{document}